\documentstyle[preprint,aps,epsf]{revtex}

\newcommand{\be}{\begin{eqnarray}}
\newcommand{\ee}{\end{eqnarray}}

\newcommand{\dslash}{D\!\!\!\!/\,}

\begin{document}
\bibliographystyle{unsrt}
\draft
\title{The axial anomaly in QCD at finite temperature}

\author{T.~Sch\"afer}

\address{Institute for Nuclear Theory, Department of Physics,
	 University of Washington,\\ Seattle, WA 98195, USA}

\maketitle

\begin{abstract}
 
  We study flavor mixing and the $U(1)_A$ anomaly in QCD at zero
and finite temperature. Using the instanton liquid model, we show 
that the strength of the anomaly is essentially unchanged near 
the critical temperature for chiral symmetry restoration. We 
demonstrate that nevertheless chiral symmetry restoration has
important consequences for the $\eta$ and $\eta'$. In particular,
the strange and non-strange components of the $\eta$ unmix near
$T_c$. The anomaly does not affect the strange eta, so we
expect a light purely strange pseudoscalar near the phase 
transition. 

\end{abstract}
\pacs{11.30.Rd, 12.38.Lg, 12.38.Mh}

  1. In connection with the ongoing heavy ion program at AGS and CERN
it is of great interest to identify possible changes in hadronic 
properties as matter is heated up and reaches the critical temperature 
for chiral symmetry restoration. Recently, a number of authors 
have argued that such changes might be very dramatic in the $\eta-\eta'$ 
sector \cite{Shu_94,KKL_96,HW_96,Coh_96} (see \cite{PW_84,HHK_88,KA_88,Kun_91} 
for earlier work on the subject). At zero temperature, the $\eta'-\pi$ 
mass splitting, which is (mostly) due to the anomaly, is larger than 
any other mesonic mass splitting. This means that any tendency towards 
(partial) $U(1)_A$ restoration might lead to physical effects that are 
more easily observed than changes in, for example, the $\rho$ meson or 
nucleon channels.  

  In this letter, we wish to study the axial anomaly and the $\eta-
\eta'$ system at finite temperature in the instanton liquid model
\cite{Shu_82a,Shu_88,SS_96}. Since instantons provide the mechanism
for the $U(1)_A$ anomaly in QCD, the applicability of the model 
appears obvious. Nevertheless, we would like to make two additional 
comments. First, the model not only accounts for the anomaly, it 
gives a very successful description of hadronic phenomenology in
general. In addition to that, the model describes spontaneous chiral 
symmetry breaking as well as its restoration at a critical temperature 
$T_c\simeq 140$ MeV. Our second comment concerns the anomaly. A 
quantitative description of the $\eta-\eta'$ system in the instanton 
model is more subtle than one might expect. Indeed, the simplest, random, 
instanton model fails to give an acceptable description of the 
$\eta'$ mass and $\eta-\eta'$ mixing. 

  The plan of this letter is as follows. First, we discuss the $\eta-
\eta'$ system at zero temperature. Next, we review general arguments
concerning the anomaly at finite temperature. Finally, we study 
$\eta$ and $\eta'$ correlation functions at finite temperature.

  2. Before we go into detail, we should remind the reader of
't Hooft's mechanism for $U(1)_A$ breaking in QCD \cite{tHo_76}.
The QCD partition function receives contributions from special
field configurations, instantons, that carry topological charge.
In the field of an instanton, the Dirac operator has a chiral zero 
mode, $i\dslash\phi_0=0$. In the case of an instanton, the zero
mode is left handed $\gamma_5\phi_0=-\phi_0$, while in the case of 
an anti-instanton it is right handed. The presence of a zero mode implies 
that for massless fermions, the amplitude for an isolated instanton 
vanishes, because it contains a factor $\det(i\dslash)$. However, 
when we calculate a $U(1)_A$ violating observable, like the 
expectation value of the 't Hooft operator ${\cal O}_{det}=\det_f(
\bar\psi_L\psi_R)+(L\leftrightarrow R)$, the $N_f$ zero modes in 
the determinant cancel against $N_f$ zero modes in the quark 
propagators. As a result, the instanton can absorb $N_f$ left 
handed quarks and turn them into right handed quarks, violating 
axial charge by $2N_f$ units.  

  In order to study the effect of the anomaly on the spectrum
of pseudoscalar mesons, we have to consider correlation functions of
the $SU(3)$ singlet and octet pseudoscalar meson currents $j_{0,8}
=\bar q\gamma_5\lambda_{0,8}q$. The diagonal singlet and octet, 
as well as the off-diagonal singlet-octet mixing correlators are
given by  
\be
\label{cor_00}
\Pi_{00} &=& \;\;\frac{1}{\,3\,}\;\; \left\{ 2\Pi^u_{con} + \;\Pi^s_{con}
 - 4\Pi^u_{dis} - \;\Pi^s_{dis} - 4 \Pi^{us}_{dis} \right\}, \\
\Pi_{88} &=& \;\;\frac{1}{\,6\,}\;\; \left\{ 2\Pi^u_{con} + 4\Pi^s_{con}
 - 4\Pi^u_{dis} - 4\Pi^s_{dis} - 8 \Pi^{us}_{dis} \right\}, \\
\label{cor_08}
\Pi_{08} &=& \frac{1}{3\sqrt{2}} \left\{ 2\Pi^u_{con} - 2\Pi^s_{con}
 - 4\Pi^u_{dis} + 2 \Pi^s_{dis} + 2 \Pi^{us}_{dis} \right\},
\ee
where $\Pi^f_{con}=\left\langle {\rm Tr}\left[S^f(x,y)\gamma_5S^f(y,x)
\gamma_5\right]\right\rangle$ is a connected correlation function and
$\Pi^{fg}_{dis}=\left\langle {\rm Tr}\left[S^f(x,x)\gamma_5\right]
{\rm Tr}\left[S^g(y,y)\gamma_5\right]\right\rangle$ is a disconnected 
correlator. Here, $S^f(x,y)$ is the quark propagator of a quark with 
flavor $f$ and $\langle .\rangle$ denotes averaging over all gauge 
field configurations. In deriving (\ref{cor_00}-\ref{cor_08}) we have 
assumed exact isospin symmetry. For comparison, the pion correlator 
is given by $\Pi_\pi=\Pi^u_{con}$ and the kaon correlator by $\Pi_K
=\Pi^{us}_{con}$. Also, the correlators of the strange and non-strange
components of the $\eta$ are given by $\Pi_{NS}=2\Pi^u_{con}-\Pi^u_{dis}$
and $\Pi_S=\Pi^s_{con}-\Pi^s_{dis}$, while their mixing is determined
by $\Pi_{NS,S}=\sqrt{2}\Pi^{us}_{dis}$. At zero temperature, we will 
exclusively focus on euclidean space correlation functions. This 
means that the long time behavior of the correlators is given by 
$\Pi\sim\exp(-m x)$, where $m$ is the ground state mass in the given 
channel. 

  Due to flavor $SU(3)$ symmetry breaking, the singlet and octet 
correlation functions mix. The physical $\eta$ and $\eta'$ states 
couple to linear combinations of the singlet and octet currents, 
\be
 \eta' &=& \cos\theta\eta_0 + \sin\theta\eta_8.
\ee 
Experimentally, the value of the mixing angle is $\theta\simeq 
-(10-20)^\circ$. The sign of the mixing angle corresponds to
a reduction of the strange component of the $\eta$ and a strangeness
enhancement in the $\eta'$. The uncertainty in the mixing angle is 
hard to assess. In fact, the concept of a mixing angle may not be 
very well defined, since the two states are rather far apart in mass.

  Qualitatively, instanton effects can be understood from the 
propagator in the field of a single instanton, $S^f(x,y)=\phi_0(x)
\phi_0^\dagger(y)/m_f + S_{NZM}(x,y)+S_m(x,y)$. Here, $\phi_0(x)$ is 
the zero mode wave function, $S_{NZM}(x,y)$ is the non-zero mode 
propagator and $S_m(x,y)$ includes mass corrections \cite{SV_93a}. 
Let us first consider the case of exact $SU(3)$ symmetry. It is 
important to note that zero modes give identical contributions to 
the connected and disconnected correlators. We find $\Pi_{00}=-2
\cdot I$, $\Pi_{08}=I$ and $\Pi_{08}=0$, where we have defined
\be
\label{zm_cont}
 I&=& \int d\rho\,\frac{n(\rho)}{m^2}\int d^4z |\phi(x-z)|^2   
 |\phi(y-z)|^2 .
\ee
Here, $n(\rho)$ is the single instanton density \cite{tHo_76b}
and $z$ is the instanton position. Note that $I$ is not singular
in the chiral limit $m\to 0$, because the instanton density  
contains a factor $m^2$. Strictly speaking, $n(\rho)$ is 
proportional to $m^3$, but if chiral symmetry is broken, 
zero modes can be absorbed by the quark condensate. This means 
that we can replace one power of $m$ by the effective mass
$m^*=\frac{4}{3}\pi\rho^2\langle\bar qq\rangle$ \cite{Shu_82a}, 
which is finite in the chiral limit. 

  

   If we include $SU(3)$ flavor breaking, the situation becomes 
more complicated. In effective models, it is usually assumed that
the 't Hooft interaction is $SU(3)$ symmetric and the mixing is
caused by quark or meson mass terms. In that case it is clear 
that the mixing angle is negative, since the mass term drives
the system toward ideal mixing, corresponding to $\theta=-54.7^\circ$.
In general, the situation is more complicated and $SU(3)$ flavor
breaking in the 't Hooft interaction is substantial. In the single
instanton approximation, flavor symmetry breaking can be taken 
into account by replacing the effective mass $m^*$ by $m^*+m_f$
\cite{Shu_83}. The off-diagonal correlators are completely 
determined by disconnected contributions, since the zero mode 
contributions to the connected and disconnected correlators 
with the same flavor cancel each other. We find \cite{Shu_83}
\be
\label{mix_sia}
 \Pi_{08}\;=\; \frac{\sqrt{2}}{3}\frac{m_u-m_s}{m^*+m_s}\,I^*,
 \hspace{1cm}
 \Pi_{NS,S}\;=\; \sqrt{2}\,\frac{m^*+m_d}{m^*+m_s}\,I^*,
\ee
where we have replaced the current quark masses by effective 
masses in $I^*$. In the single instanton approximation $\Pi_{08}$ 
is negative, which corresponds to a positive mixing angle\footnote{
The off-diagonal correlator in the single instanton approximation
was first calculated in \protect\cite{Shu_83}, but the 
conclusions concerning the mixing angle appear to be wrong.}. 
On the other hand, mass corrections in the non-zero mode part 
of the propagator give a positive contribution to $\Pi_{08}$. 
This means that the sign of the mixing angle is determined by 
the competition between the flavor breaking in direct mass 
insertions and the 't Hooft interaction. 

 In the following, we study this problem in the instanton liquid 
model. We refer the reader to \cite{SS_96,SS_96b} for details of 
the model. For our purposes here it is only important that there
are three different instanton ensembles, the random (RILM), quenched
(QILM) and unquenched (IILM) models. In the random and quenched 
ensembles, the topological susceptibility is finite $\chi_{top}
\simeq (N/V)$, where $(N/V)$ is the density of instantons. In 
the unquenched ensemble, topological charge is screened and
$\chi_{top}\sim m\langle\bar qq\rangle$. Eta meson correlation 
functions in the random model\footnote{The results in the quenched 
ensemble are very similar. Eta meson correlation functions in the
random ensemble were first studied in \cite{SV_93b}. Unfortunately,
this work contains an error in the flavor octet and the off-diagonal
singlet-octet correlation functions. In particular, $\Pi_{08}$ has 
the wrong sign.}  and two different unquenched 
ensembles (the ``stream line" and ``ratio ansatz" ensembles, see 
\cite{SS_96}) are shown in figure \ref{fig_eta}. We note that the 
$U(1)_A$ anomaly is ``over-explained" in the random ensemble. The 
$\eta'$ channel is so repulsive that the correlation function becomes 
unphysical ($\Pi(x)<0$). Also, flavor symmetry breaking is very strong 
and the singlet-octet correlator is large and negative (corresponding 
to a positive mixing angle). 

  This situation is improved in the unquenched ensembles. The 
flavor-singlet correlation function is physical and singlet-octet
mixing is smaller. The prediction for the masses and mixing angles
depends sensitively on details of the interaction. In the streamline
ensemble, the $\eta'$ is still too heavy, $m_{\eta'}\simeq 2$ GeV. 
The mass of the $\eta$ is given by $m_\eta=(0.66\pm 0.12)$ GeV 
and the mixing angle is small, $\theta=(1\pm 3)^\circ$. A small
mixing angle was also found in the Nambu and Jona-Lasinio model
\cite{TO_95} (see also \cite{HK_91,KLV_90}). In particular, these 
authors showed that a small mixing angle is not necessarily incompatible 
with the observed $\eta\to 2\gamma$ rate.  

  3. A global measure of the strength of the anomaly is provided 
by the expectation value of the 't Hooft operator. In figure 
\ref{fig_4q} we show the temperature dependence of $\langle{\cal O}_{det}
\rangle$ in the finite temperature instanton ensembles\footnote{
All of these ensembles have total topological charge $Q=0$. This 
means that the topological susceptibility evaluated for the entire
volume is not correct. This should not affect local observables.
Indeed, we have checked that topological charge fluctuations in
a sub-volume have the expected dependence on the quark mass and
volume \protect\cite{SV_95}.} obtained 
in \cite{SS_96}. For comparison, we also show the quark condensate 
and two different four fermion operators. At zero temperature both the
$SU(2)\times SU(2)$ chiral and the axial $U(1)_A$ symmetry are 
broken. Chiral symmetry breaking is caused by interactions between
the zero modes associated with individual instantons. As a result,
some of the lowest states become collective and form a condensate.
Near $T=125$ MeV, chiral symmetry is restored and the quark condensate 
goes to zero. This transition is due to a rearrangement of the 
instanton liquid, going from a disordered, random, system to an
ensemble of topologically neutral instanton-anti-instanton pairs. 
Clearly, there is no tendency towards $U(1)_A$ restoration as chiral 
symmetry is restored. In fact, $\langle{\cal O}_{det}\rangle$ has a 
maximum near the phase transition (although the uncertainty is also 
largest near $T_c$). The reason why $\langle{\cal O}_{det}\rangle$ 
survives the chiral phase transition should be clear from our 
discussion above: the 't Hooft operator can induce a tunneling 
event all by itself\footnote{This can also be verified in the mean
field approximation to the instanton liquid for two flavors,
see \protect\cite{SV_96}}. At temperatures significantly above $T_c$ 
the semi-classical tunneling amplitude contains the suppression 
factor $n(\rho)\sim \exp(-(2N_c/3+N_f/3)(\pi\rho T)^2)$ \cite{GPY_81} 
and $\langle{\cal O}_{det} \rangle$ becomes small. This suppression 
factor is mostly due to Debye screening of the instanton field. 
Therefore, it does not affect the instanton density below the 
phase transition. This was checked explicitly by performing a 
soft pion calculation of the instanton density at small temperature 
\cite{SV_94} and in (quenched) lattice calculations of the instanton 
density at finite temperature \cite{CS_95,IMM_95,ADD_96}.

  Local four fermion operators like ${\cal O}_{det}$ are hard to
measure on the lattice. Therefore, a number of authors have studied 
$U_A(1)$ violating mesonic susceptibilities. Here, mesonic 
susceptibilities are defined as integrals of the corresponding 
correlation function, $\chi_\Gamma =\int d^4x\,\Pi_\Gamma(x)$. 
The most natural candidate for a $U(1)_A$ order parameter is 
the difference $\chi_\pi-\chi_{\eta'}$, but this quantity also 
involves disconnected quark loops. A better observable is $\chi_\pi-
\chi_\delta$, originally suggested by the Columbia group \cite{Cha_95}. 
For $N_f=2$ and if chiral symmetry is restored, this quantity is a 
measure of $U(1)_A$ breaking. A nice feature of $\chi_\pi-\chi_\delta$ 
is that it can be expressed in terms of the spectral density 
$\rho(\lambda)$ of the Dirac operator
\be
 \chi_\pi-\chi_\delta = 4m^2 \int d\lambda
   \frac{\rho(\lambda)}{(\lambda^2+m^2)^2}.
\ee
For comparison, the quark condensate is given by 
\be
 \langle\bar qq\rangle =-2m \int d\lambda
   \frac{\rho(\lambda)}{\lambda^2+m^2}.
\ee
These results allow us to constrain the low virtuality part 
of the spectrum. For $U(1)_A$ to be broken but chiral symmetry
restored, we require $\chi_\pi-\chi_\delta$ to be finite in the 
limit $m\to 0$ while $\langle \bar qq\rangle$ goes to zero. 
This requirement is clearly satisfied 
by $\rho(\lambda)\sim m^{N_f}\delta(\lambda)$, corresponding 
to a dilute system of instantons. However, interactions among 
zero and non-zero modes might alter the shape of the spectrum. 
It is therefore interesting to note that the criterion given 
above is also satisfied by a non-analytic spectral density 
$\rho(\lambda)\sim \lambda^\alpha$ with $\alpha\leq 1$.

  In order to study this question in more detail, we have 
determined the spectrum of the Dirac operator in the instanton
liquid (for $N_f=2$) for several different values of the 
temperature and the quark masses, see figure \ref{fig_dirac}. 
Above $T_c$ we clearly observe a peak in the spectrum near
$\lambda=0$. The number of eigenvalues in the peak is nicely
consistent with $N(\lambda\simeq 0)\sim m^2$. Below $T_c$, 
most of the small eigenvalues are related to chiral symmetry 
breaking. Their number is proportional to the effective mass
$m^*$ and almost independent of the current mass $m$. There 
is a dip in the spectrum for small quark masses. This is a
finite volume effect. In a finite volume, the spectral 
density near zero will always go to zero as $m\to 0$. We have
checked that the width of the dip in the spectrum decreases 
as the volume is increased. 

  A number of groups have measured $\chi_\pi-\chi_\delta$
(or the corresponding screening masses) on the lattice
\cite{Cha_95,Boy_96,Ber_96}. Most of the published results indicate
that $U(1)_A$ remains broken, although recent results by
the Columbia group have questioned that conclusion \cite{Chr_96}. 
From the literature, it is not clear whether lattice simulations
find the peak in the spectrum observed in the instanton 
liquid. The Columbia group has measured the valence mass 
dependence of the quark condensate, which is a folded 
version of the Dirac spectrum \cite{Cha_95}. Their result
looks very smooth, not indicative of a small virtuality 
peak. One should note, however, that instanton calculations
focus exclusively on the small virtuality part of the spectrum
while the number of eigenvalues in lattice simulations is much 
larger. Also, both instanton and lattice simulations suffer 
from certain artefacts if the quark mass is made too small.  
In our case, if the mass is too small, isolated instantons
are rare and the constraint $Q=0$ affects the results. On 
the lattice, for small quark masses one might run into
problems with chiral fermions. 

  4. The $U(1)_A$ anomaly at finite temperature is usually 
discussed in terms of the effective lagrangian \cite{PW_84}
\be
\label{l_eff}
 {\cal L} &=& \frac{1}{2}{\rm Tr}\left((\partial_\mu\Phi)
  (\partial_\mu\Phi^\dagger)\right) - {\rm Tr}\left({\cal M}
  (\Phi+\Phi^\dagger)\right) + V(\Phi\Phi^\dagger)
  + c \left(\det\Phi+\det\Phi^\dagger\right),
\ee
where $\Phi$ is a meson field in the $(3,3)$ representation of 
$U(3)\times U(3)$, $V(\Phi\Phi^\dagger)$ is a $U(3)\times U(3)$
symmetric potential (usually taken be quartic), ${\cal M}$ is a 
mass matrix and $c$ controls the strength of the $U(1)_A$ breaking
interaction. If the coupling is taken to be $c=\chi_{top}/(12f_\pi^3)$,
the effective lagrangian reproduces the Witten-Veneziano relation 
$f_\pi^2m_{\eta'}^2=\chi_{top}$. In a quenched ensemble, we can further
identify $\chi_{top}\simeq (N/V)$. The temperature dependence
of $c$ is usually estimated from the semi-classical tunneling
amplitude $n(\rho)\sim \exp(-(8/3)(\pi\rho T)^2)$. As a result, 
the strength of the anomaly is reduced by a factor $\sim 5$ 
near $T_c$. If the anomaly becomes weaker, the eigenstates are 
determined by the mass matrix. In that case, the mixing angle is 
close to ideal $\theta=-54.7^\circ$ and the non-strange $\eta$ is 
almost degenerate with the pion. 

   There are several points in this line of argument that are not 
entirely correct. The strength of the 't Hooft term is not 
controlled by the topological susceptibility ($\chi_{top}=0$
in full QCD!), $\chi_{top}$ is not proportional to the instanton
density (for the same reason), and, at least below $T_c$, the
semi-classical estimate for the instanton density is not 
applicable. As we saw above, at $T_c$ the instanton liquid 
is rearranged but the strength of the $U(1)_A$ anomaly does 
not change very much. However, chiral symmetry restoration 
affects the structure of flavor mixing in the $\eta-\eta'$ 
system (see figure \ref{fig_mix}). The mixing between the
strange and non-strange eta is controlled by the light
quark condensate (see equation (\ref{mix_sia})), so 
$\eta_{NS}$ and $\eta_S$ do not
mix above $T_c$. As a result, the mixing angle is not
close to zero, as it is at $T=0$, but close to ideal. 
Furthermore, the anomaly can only affect the non-strange 
$\eta$, not the strange one. Therefore, if the anomaly is 
sufficiently strong, the $\eta_{NS}$ will be {\em heavier} 
than the $\eta_S$.

   We should compare this scenario to other possibilities
discussed in the recent literature. A number of authors
have noticed that for three massless flavors, both the 
$\eta$ and $\eta'$ are unaffected by the anomaly above 
$T_c$ \cite{SS_96b,EHS_96,LH_96,BCG_96}. This is a related
effect, but not that relevant for QCD, where the strange 
mass is not small. Also, two recent papers have argued
that the $\eta_{NS}$ and $\eta_S$ decouple near $T_c$
\cite{KKL_96,HW_96}. However, in these works the effect
is caused by the disappearance of the anomaly and as 
a result, the $\eta_S$ is always predicted to be heavier
than the $\eta_{NS}$. The scenario proposed here is 
consistent the effective lagrangian\footnote{
Our scenario cannot be described in terms of the non-linear
effective lagrangian employed in \cite{HW_96}. This should
not be suprising; non-linear effective meson theories have
to be used with care near the chiral phase transition.} 
(\ref{l_eff}). However, most authors employ a ``first 
order" treatment of flavor symmetry breaking and neglect
terms of order $(m_s-m_u)c$. These terms are precisely 
what causes the effect discussed here.  

   5. To explore this phenomenon in a more quantitative way,
we study $\eta-\eta'$ correlation functions in the instanton 
model at finite temperature. The results were obtained
for three flavors with masses $m_u=m_d=22$ MeV and $m_s=
155$ MeV. We consider temporal correlation functions, 
rather than the spacelike screening correlators usually
calculated on the lattice. Temporal correlators have the 
advantage that they are directly related to the spectrum 
of physical excitations. 

  The results are shown in figure \ref{fig_teta}. Correlation 
functions below $T_c$ are shown by the open squares ($T=55$ MeV), 
pentagons ($T=78$ MeV) and hexagons ($T=111$ MeV), while the 
correlators near and above $T_c$ are denoted by closed squares 
($T=126$ MeV) and pentagons ($T=145$ MeV). Below $T_c$ the
singlet correlation function is strongly repulsive, while
the octet correlator shows some attraction at larger distance.
The off-diagonal correlator is small and positive, corresponding
to a negative mixing angle. The strange and non-strange eta
correlation functions are very similar, which is a sign for
strong flavor mixing. This is also seen directly from  
the off-diagonal correlator between $\eta_S$ and $\eta_{NS}$.

   Above $T_c$, the picture changes. The off-diagonal singlet-octet 
changes sign and its value at intermediate distances $\tau\simeq   
0.5$ fm is significantly larger. The strange and non-strange
eta correlators are very different from each other. The non-strange
correlation function is very repulsive, while the strange one
is significantly larger. This clearly supports the scenario
presented above. Near $T_c$ the eigenstates are essentially
the strange and non-strange components of the $\eta$, with the
$\eta_S$ being the lighter of the two states. This picture 
is not realized completely, $\Pi_{S,NS}$ does not vanish and
the singlet eta is still somewhat more repulsive than the 
octet eta correlation function. This is due to the fact that
the light quark mass does not vanish. In particular, in this 
simulation the ratio $(m_u+m_d)/(2m_s)=1/7$, which is about 
three times larger than the physical mass ratio.

   It is difficult to provide a quantitative analysis of temporal   
correlation functions in the vicinity of the phase transition. At 
high temperature the temporal direction in a euclidean box becomes 
short and there is no unique way to separate out the contribution
from excited states. Nevertheless, under some simplifying assumptions
one can try to translate the correlation functions shown in figure
\ref{fig_teta} into definite predictions concerning the masses 
of the $\eta$ and $\eta'$. For definiteness, we will use ideal
mixing above $T_c$ and fix the threshold for the perturbative 
continuum at 1 GeV. In this case, the masses of the strange
and non-strange components of the $\eta$ at $T=126$ MeV are 
given by $m_{\eta_S}=(0.420\pm 0.120)$ GeV and $m_{\eta_S}=
1.250\pm 0.400)$ GeV.

6. In summary, we studied flavor mixing and the axial anomaly 
at zero and finite temperature. At zero temperature, we have emphasized
two points. The first one is that the correlations among instantons
that lead to topological charge screening are also important in
reproducing the $\eta$ and $\eta'$ correlation functions. The 
second one is that flavor symmetry breaking in the instanton 
induced interaction is substantial. It acts against the symmetry
breaking from direct mass insertions. As a result, the $\eta-\eta'$ 
mixing angle is small. 

  At finite temperature, we have shown that the strength of the
anomaly as measured by the expectation value of the 't Hooft 
operator is essentially independent of temperature below $T_c$.
In our model, $\langle {\cal O}_{det}\rangle$ even peaks near
$T_c$ and then drops at larger temperatures. In terms of the 
spectral density of the dirac operator, the anomaly is due to
a spike $\rho(\lambda)\sim m^{N_f}\delta(\lambda)$ in the 
spectrum at zero virtuality. 

   Our most important result concerns the role of flavor mixing 
at finite temperature. Although the strength of the anomaly is 
not reduced near $T_c$, chiral symmetry restoration affects the 
$\eta-\eta'$ system. If the light quark condensate vanishes, 
transitions between light and strange pseudoscalars are suppressed. 
As a result, the eigenstates are given by the strange and 
non-strange components of the $\eta$. The anomaly does not
affect the strange $\eta$, so we predict a purely strange,
light pseudoscalar near the transition. We have estimated the
mass of this state to be around 400 MeV. A light strange 
meson is of interest in connection with strangeness production 
in relativistic heavy ion collisions. This suggests that 
the coupling of a light $\eta_S$ to kaons and etas should 
be studied in more detail. 

Acknowledgements: I would like to thank V. Koch and E. Shuryak
for useful discussions.



\newpage\noindent
\begin{figure}
\caption{\label{fig_eta}
Singlet and octet eta correlation functions in different
instanton ensembles at zero temperature. All correlation
functions are normalized to free quark propagation.}
\end{figure} 
\begin{figure}
\caption{\label{fig_4q}
Temperature dependence of the quark condensate, the 't Hooft 
operator and two different four fermion operators in the 
instanton liquid. All condensates are normalized to their
$T=0$ values.}
\end{figure} 
\begin{figure}
\caption{\label{fig_dirac}
Dirac spectra below and above the chiral phase transition for 
different dynamical quark masses. The spectral density is given
in arbitrary units. The instanton density was held fixed at
$(N/V)\Lambda^{-4}=1$, where $\Lambda$ is the QCD scale parameter.
Quark masses and inverse temperature $\beta=T^{-1}$ are  
given in units of $\Lambda$.}
\end{figure} 
\begin{figure}
\caption{\label{fig_mix}
Leading contributions to flavor mixing in the $\eta-\eta'$
system below and above the chiral phase transition.}
\end{figure} 
\begin{figure}
\caption{\label{fig_teta}
Correlation functions for singlet and octet, off-diagonal 
singlet-octet as well as strange and non-strange etas at 
different temperatures. Open squares, pentagons and hexagons
are below $T_c$, closed squares and pentagons above $T_c$. }
\end{figure} 

\setcounter{figure}{0}
\newpage

\pagestyle{empty}

\newpage
\begin{figure}
\begin{center}
\leavevmode
\epsfxsize=16cm
\vspace*{-3cm}
\epsffile{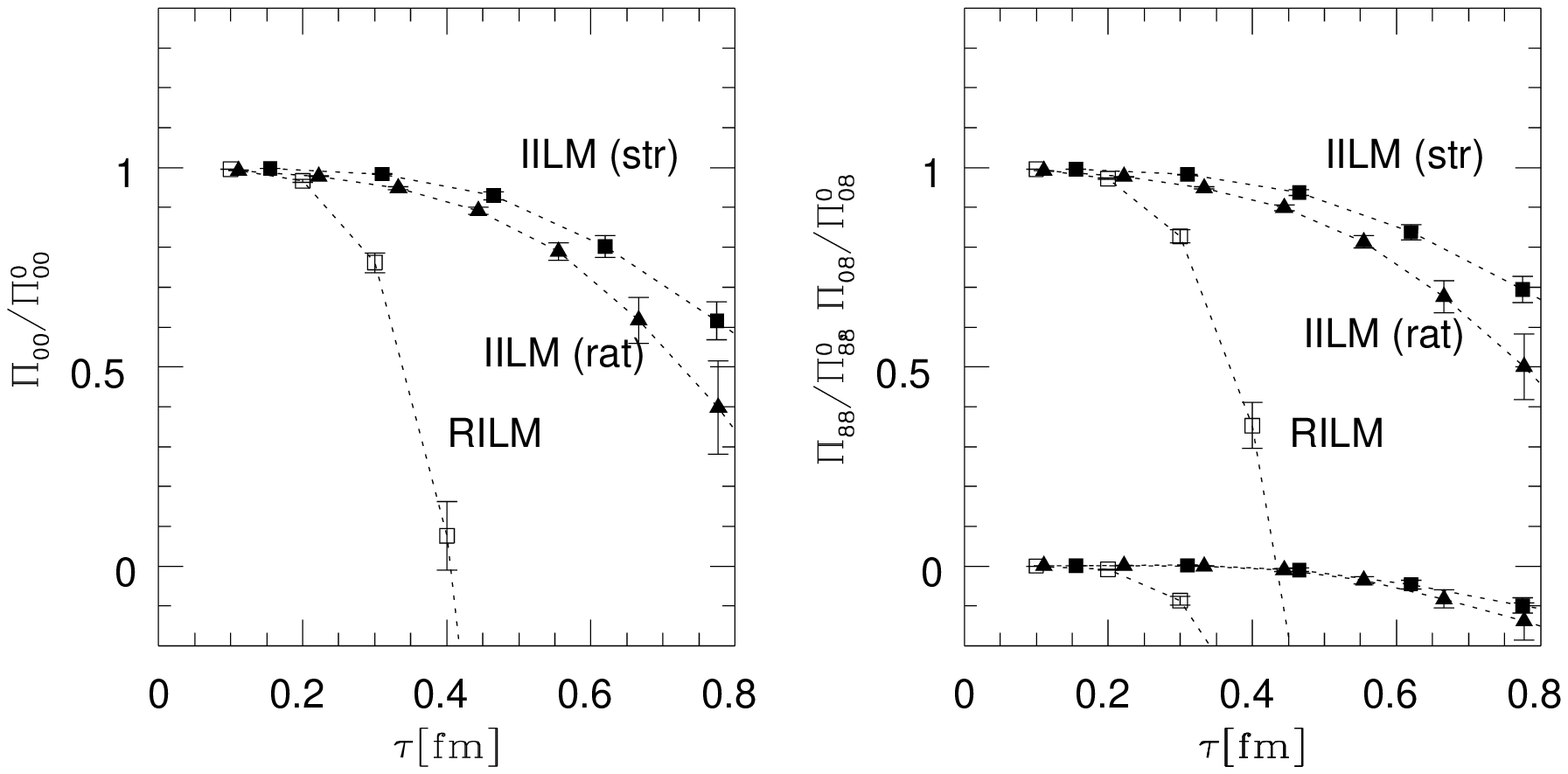}
\end{center}
\vspace*{-1cm}
\caption{}
\end{figure}

\begin{figure}
\begin{center}
\leavevmode
\epsfxsize=8cm
\epsffile{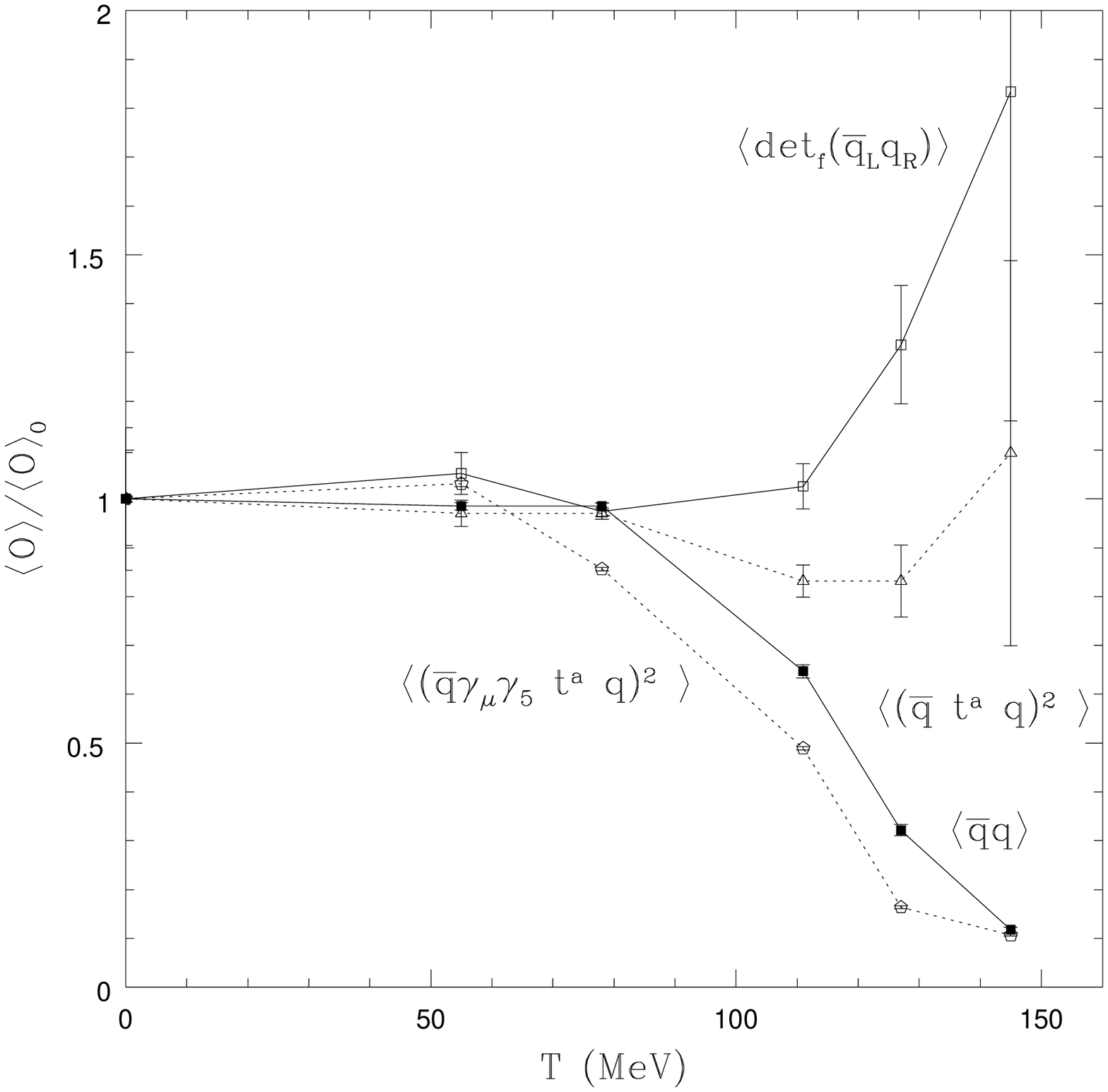}
\end{center}
\caption{}
\end{figure}
\vfill

\begin{figure}
\begin{center}
\leavevmode
\epsfxsize=16cm
\epsffile{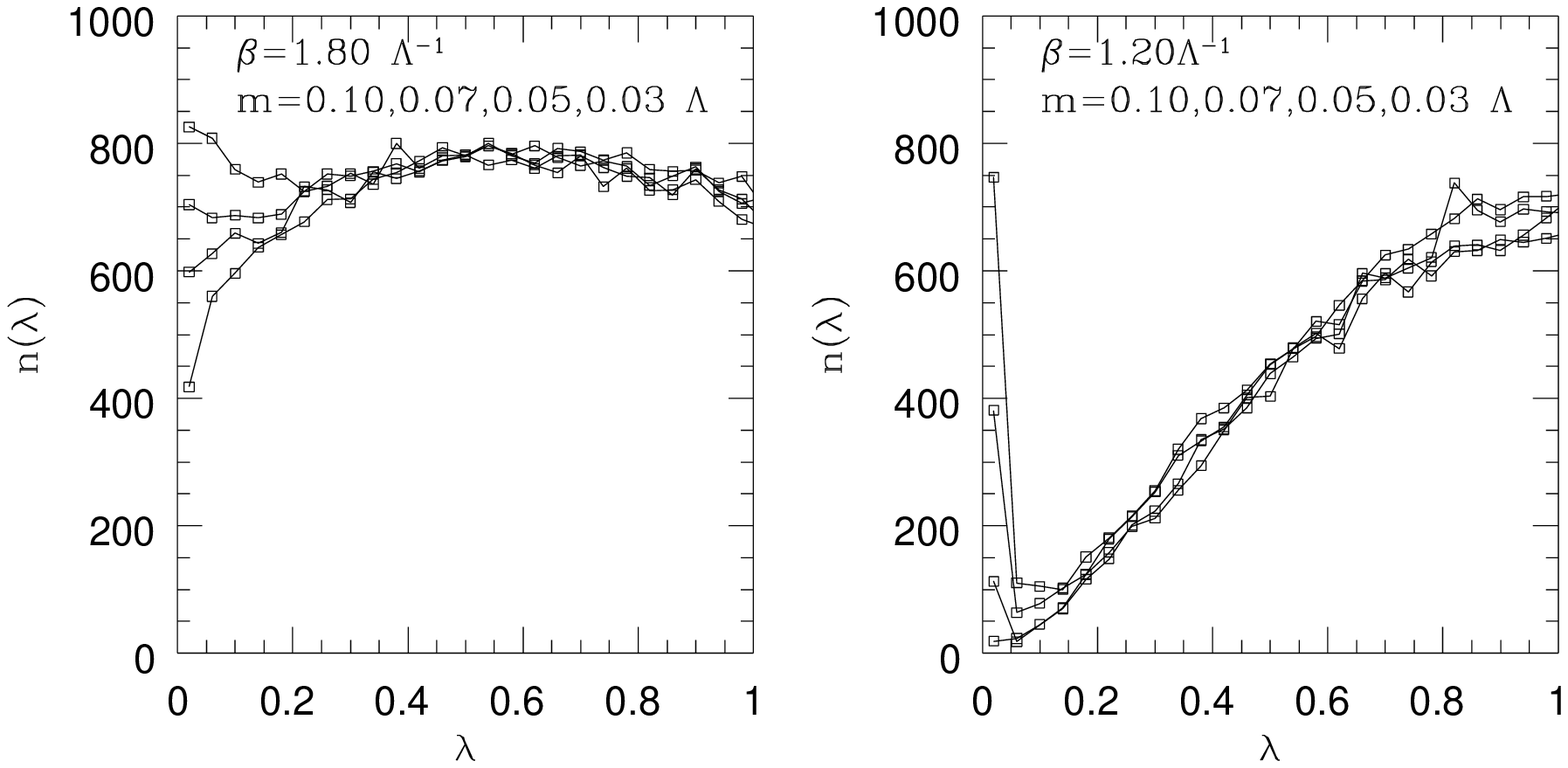}
\end{center}
\vspace*{-3cm}
\caption{}
\end{figure}
\vfill

\begin{figure}
\begin{center}
\leavevmode
\epsfxsize=16cm
\epsffile{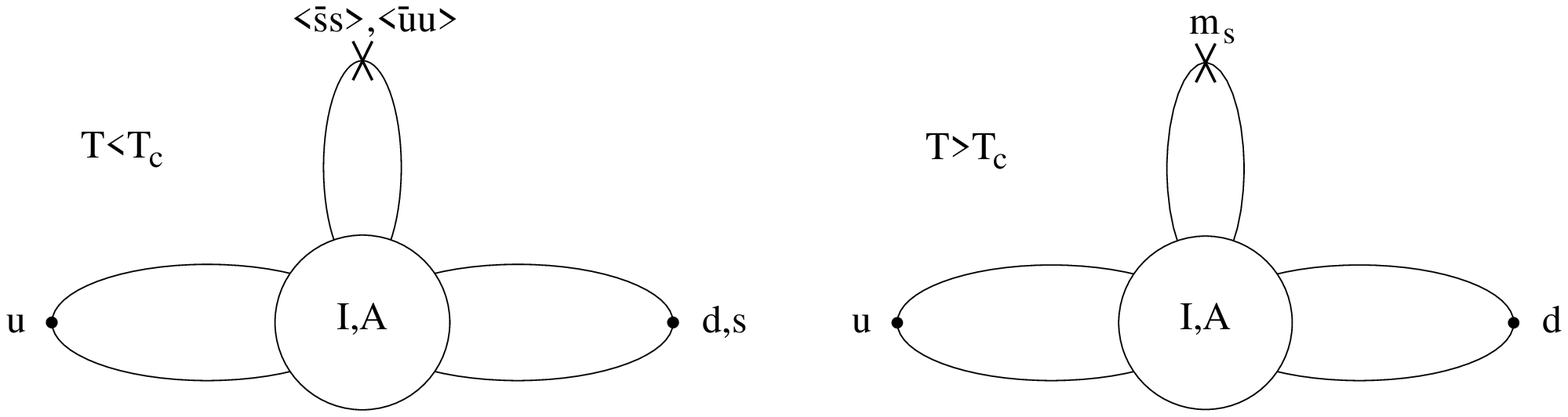}
\end{center}
\caption{}
\end{figure}
\vfill

\begin{figure}
\begin{center}
\leavevmode
\epsfxsize=16cm
\epsffile{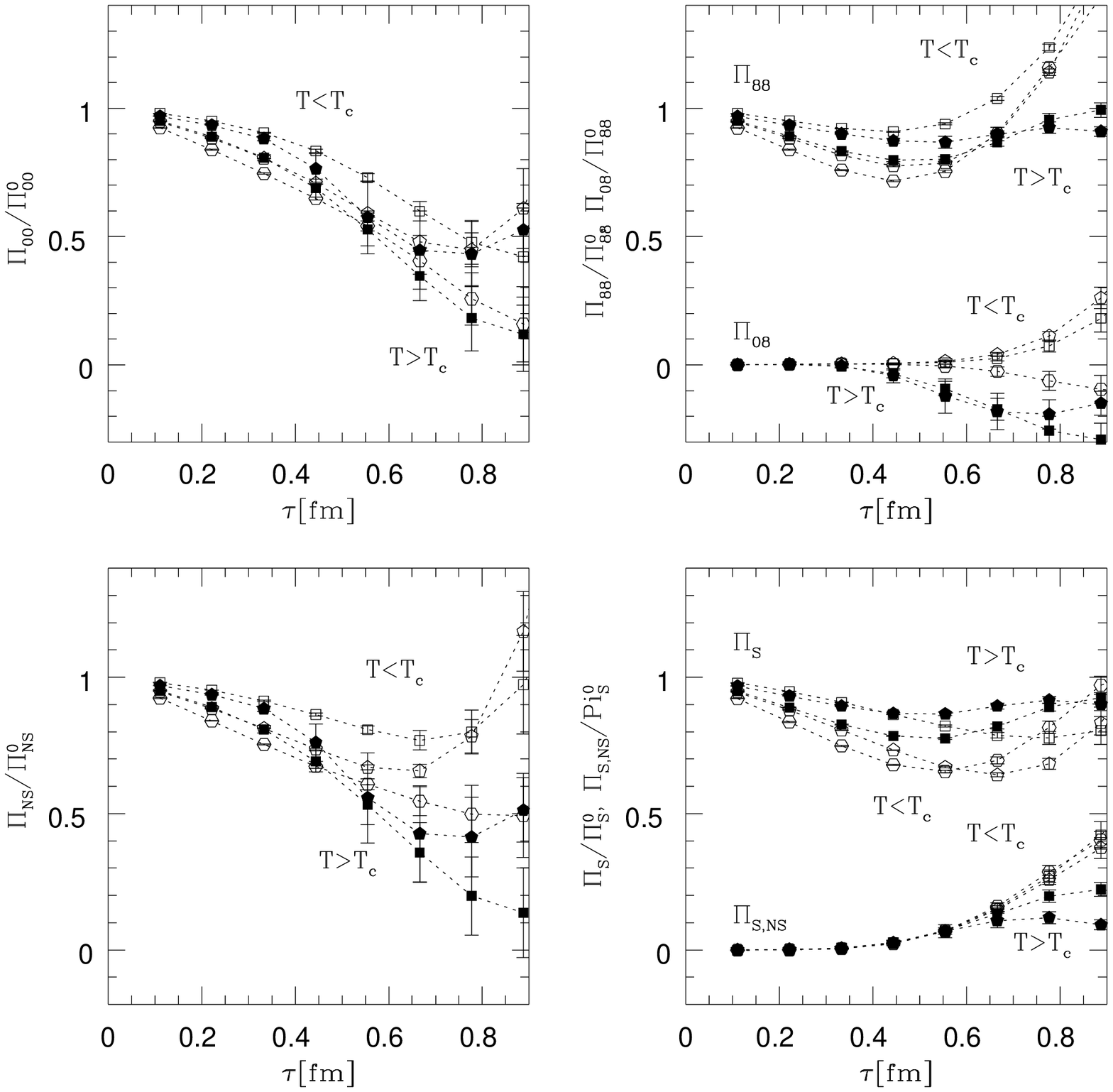}
\end{center}
\caption{}
\end{figure}
\vfill

\end{document}